
\magnification=1200
\pretolerance=1000
\parindent 5pt
\parskip=0.5cm
\hsize 12.5cm
\baselineskip=0.5cm

\line{\hfil {\bf BINARY TREE APPROACH TO SCALING IN UNIMODAL MAPS}\hfil}
\vskip 1cm
\line{\hfil Jukka A. Ketoja\hfil}
\medskip
\line{\hfil Research Institute for Theoretical Physics,
P.O.Box 9,\hfil}
\line{\hfil SF-00014 University of Helsinki, Finland\hfil}
\vskip 1cm
\line{\hfil Juhani Kurkij\"arvi\hfil}
\medskip
\line{\hfil Institutionen f\"or fysik, {\AA}bo Akademi,\hfil}
\line{\hfil Porthansgatan 3, SF-20500 {\AA}bo, Finland\hfil}
\vskip 2cm
{\centerline{\bf Abstract}}
Ge, Rusjan, and Zweifel (J. Stat. Phys. 59, 1265 (1990)) introduced
a binary tree which represents all the periodic windows in the
chaotic regime of iterated one-dimensional unimodal maps. We consider
the scaling behavior in a modified tree which takes into account
the self-similarity of the window structure. A non-universal
geometric convergence of the associated superstable parameter
values towards a Misiurewicz point is observed for almost all
binary sequences with periodic tails. There are an
infinite number of exceptional sequences, however, which lead to
superexponential scaling. The origin of such sequences is explained.

\vfill\eject

{\bf 1. INTRODUCTION}

Iterated one-dimensional unimodal maps [1] have been the subject
of hundreds of papers over the last couple of decades.
Not only are they among the simplest dynamical systems exhibiting
chaotic behavior but also very important as prototypes
of dissipative systems. In spite of breakthroughs
such as
symbolic dynamics [1,2,3], ergodic behavior [1,4,5],
and the transition to chaos [6] in the theory of unimodal maps,
many problems remain related
to the scaling behavior of such maps within
the so called chaotic regime.
This region is defined as the parameter
interval between the first period-doubling accumulation point
and the final crisis point beyond which no periodic or chaotic
attractors can be found within the unimodality interval of the
phase space.

Rigorous mathematical proofs [5] establish that
the parameter values corresponding to an absolutely continuous
invariant ergodic measure form a set with a positive Lebesgue measure.
These "chaotic" parameter values are found
in between the infinite
number of windows with
stable periodic attractors. The "periodic" windows are expected
to be dense
on the parameter axis. Although each window has a finite lenght
there remains a great deal of space for chaotic parameter values:
Near the accumulation point of a period-doubling
cascade, the relative fraction of the aperiodic solutions is
given by the universal number $0.892...$ [7]. Even considering
the whole chaotic region, the probability of finding
an aperiodic solution
is approximately $9/10$ for a typical map [8].

Since periodic windows are ubiquitous along the parameter axis,
various infinite sequences of them are a natural tool when
investigating scaling properties of unimodal maps.
The scaling behaviors of
period-doubling [6] and more general multifurcation sequences [9],
period-adding sequences approaching
a crisis [10,11] and tangent bifurcation points [12] have been
determined.  Shibayama [13] extended the analysis to
the so called Fibonacci sequences whose scaling is superexponential
both on the parameter axis and in the phase space.
An exact universal form of that type of scaling was
found by Ketoja and Piiril\"a [14]
using a renormalization argument.  Later Lyubich and Milnor [15]
derived rigorous results both for the scaling behavior
and the dynamics at the accumulation points of such sequences.

In this paper a multitude of new examples of both
non-universal geometric and universal superexponential scaling are
reported and
the corresponding sequences related to
a binary tree of periodic windows introduced by
Ge, Rusjan, and Zweifel (GRZ) [16].
Originally the tree was defined so that each window
included the period-doubling tail in addition to
the stable parameter interval of a periodic solution.
In our modified tree,
each window is extended up to the corresponding interior
crisis point.  We "sum" not only over the period-doubling
tail but over all the multifurcation sequences. In this way
the self-similarity of the periodic-window structure [17] can be
naturally taken into account. The structure within each window
is essentially a small copy of the entire structure along the
parameter axis. We concentrate on those periodic-window  sequences
whose binary codes have periodic tails. Such sequences usually
lead to arithmetic growth of the period and to non-universal
geometric scaling. The windows accumulate at a Misiurewicz point [1,4]
at which the dynamics is completely chaotic. In addition, we
find an infinite number of exceptional cases in which
the period increases geometrically and the scaling is superexponential
generalizing earlier results on Fibonacci sequences.
Finally, preliminary numerical conclusions on
aperiodic binary codes are reported.
It is conjectured that the accumulation
point of such a sequence always corresponds to a chaotic attractor.

{\bf 2. BINARY TREE}

Consider a one-parameter family $f_{\mu} (x)$ of differentiable
unimodal maps from a real interval $I$ to itself.
$f_{\mu}$ is assumed to have a quadratic maximum at $x=c$
so that the map is
monotonically increasing for $x\in I, \;x<c$
and monotonically decreasing on the other side of the critical
point $c$. An orbit obtained by iterating $f_{\mu}$ starting from $c$
can be symbolically represented in terms of the kneading sequence
$a_1 a_2 ...$ where $a_i =R$ if $f^i_{\mu} (c) >c$ and $a_i =L$ if
$f^i_{\mu} (c)<c$.
The case in which the orbit returns back to the critical point
after $i$ iterations, $f^i_{\mu} (c)=c$,
is indicated by cutting the kneading
sequence after $i-1$ symbols so that the sequence becomes finite.
Finite symbol sequences therefore correspond to superstable periodic
orbits. We are interested in the admissible
kneading sequences at some parameter values of the unimodal
map. Metropolis, Stein, and Stein (MSS) [2] discovered
a simple rule by which
all admissible symbol sequences, the so called MSS sequences,
can be generated and arranged on the parameter axis. Originally
the rule was developed with one-parameter families of one-dimensional
maps in mind, the type where the parameter appears as a
multiplicative factor in the definition
of the map. It holds, nevertheless, in a much larger class of unimodal
maps, e.g. the logistic map $f_{\mu} (x)=\mu -x^2$.

It is instructive to take a look at the origin of the rule.
To this end, consider two parameter values
$\mu_1 $ and $\mu_2$ ($\mu_1 < \mu_2 $) which
correspond to the infinite MSS sequences $A$ and $B$.
As in ref. [16], the beginning shared by
the sequences is denoted by $A\wedge B$.
In other words, $\mu_1$ corresponds to the symbolic orbit
$(A\wedge B)a_n ...$ and $\mu_2$ to the orbit
$(A\wedge B)b_n ...$, where the symbols $a_n$ and $b_n$ differ.
The parameter dependence is assumed such that
the MSS sequences in the interval $(\mu_1 ,\mu_2 )$
always begin with $A\wedge B$. By continuity there has to be
a parameter value $\mu_3 \in (\mu_1 ,\mu_2 )$ with
the finite symbol sequence $C=A\wedge B$. The corresponding
orbit is superstable with the period $n$.  Within the interval $(\mu_1 ,
\mu_2 )$, there are no other orbits with periods $\leq n$.
Above $\mu_3$, but within the window
for the stable period $n$, the MSS sequence has the form
$h(C)=Cb_n Cb_n Cb_n ...$ (the $n$th symbol has to be $b_n$ ---
otherwise there would be at least two parameter values
with the MSS sequence $C$). From $h(C)$ and
$B$ a new periodic window in between $\mu_3$ and $\mu_2$ can
now be constructed. Similarly, the infinite
sequences $A$ and $a(C)=Ca_n Ca_n Ca_n ...$ can be chosen to
generate a periodic window in between $\mu_1$ and $\mu_3$.
The only way this recursive procedure of generating new periodic
windows can get stuck at some level is two infinite
MSS sequences' becoming equal. By repeating the procedure an
infinite number of times, MSS sequences can be generated
for aperiodic orbits as well.

One needs two MSS sequences in order to get the procedure
going.  For a so-called full
one-parameter family of unimodal maps [1], such as the
logistic map, one can set out with the symbol sequences for
the superstable period-two cycle ($R$) and the final crisis point
($RLLL...$). In the latter, $c$ is mapped onto an unstable fixed point
after two iterations.

The infinite sequences $h(C)$ and $a(C)$ above are called
the harmonic and antiharmonic extensions of $C$.
These extensions can be expressed
without knowing the infinite "parent"
sequences $A$ and $B$. Let us first write the extensions
in the form
$$\eqalign{h(C) &=C\alpha C\alpha C\alpha ...  \cr
a(C)&=C\bar \alpha C\bar \alpha C\bar \alpha ...\cr}$$
where $\bar \alpha$ denotes the "conjugate" of the symbol $\alpha$;
i.e. $\bar R=L$ and $\bar L=R$. In the sequel we will refer to $\alpha$
or $\bar \alpha$ as binding elements of an extension.
Around the critical point $c$,
there is a small box in which the $n$th iterate of
the unimodal map looks very similar to the first iterate.
The $n$th iterate has either a maximum or a minimum at $c$.
At the superstable parameter value with the MSS sequence $C$,
the extremum touches the critical point. As the value
of the parameter is increased within the stability interval, the
extremum passes either below or above the critical point
depending on whether the extremum is a maximum or a minimum.
If $C$ contains an even number of $R$'s
($C$ even) the $n$th iterate has a maximum at $c$;
otherwise ($C$ odd) the extremum
is a minimum [11]. In other words, $\alpha =R$ in the former
and $\alpha =L$ in the latter case.

The fact that the $n$th iterate restricted to a small box
around the critical point becomes a unimodal map is responsible
for the self-similarity of the MSS structure. One expects the
same MSS periodic windows with the $n$th iterate as
with the original map. Only the structure in the higher
iterate is observed in a much smaller parameter interval than for the
original map.
The window with the "reduced" MSS sequence $a_1 ... a_k$, which
includes only every $n$th iterate (these actually
land inside the small box),
corresponds to the full MSS sequence [17]
$$C*(a_1 ... a_k )
=\cases{ Ca_1 Ca_2 ...Ca_k C ,\hbox{ if }C\hbox{ is even } \cr
C\bar a_1 C\bar a_2 ...C\bar a_k C ,\hbox{ if }C\hbox{ is odd } \cr}$$
for the original map. By this composition law it is easy to write
down the MSS sequence at the endpoint of the parameter interval which
contains the self-similar copy of the whole periodic window structure:
$C*(RLLL...)=C\alpha C \bar \alpha C \bar \alpha ... =
C\alpha a(C)$. This endpoint corresponds to an internal crisis where
the orbit of
the critical point lands on an unstable period $n$ after $2n$
iterations. We call $C\alpha a(C)$ the crisis extension and
denote it by $e(C)$.

The superstable period-two cycle with the MSS code $R$, preceded
by a stable fixed point on the parameter axis, belongs to the
primary period-doubling cascade which ends at the transition
to chaos. By self-similarity, every periodic window is followed
up by a similar cascade. GRZ [16] modify
the definition of a periodic window including in it the
corresponding period-doubling tail. In the recursive
construction of the periodic windows
one considers, instead of $h(C)$, the MSS code at the
period-doubling accumulation
point, $\hat h(\hat h(\hat h(...\hat h(C)...)))$, where the
"cut" harmonic extension $\hat h(C)=C\alpha C$ is successively applied
an infinite number of times. In this way, one never generates windows
within a period-doubling cascade and
two neighboring infinite sequences never become equal. Therefore, it is
possible to generate an infinite binary tree of periodic windows.
GRZ want to
apply Feigenbaum's [18] general ideas on the renormalization of
binary trees to this particular case. We take a different point of view
and exploit the binary tree as a tool of studying the overall
scaling behavior within the chaotic region.

We  modify the GRZ tree so as to take
into account the self-similarity of the window structure.
Each window is extended up to the corresponding interior crisis point.
The infinite binary tree is constructed taking advantage of the
antiharmonic and crisis extensions in the following way:\hfil\break
1) Begin with the infinite "left parent"  $e(R)$ and the
infinite "right parent" $RLLL...$.
The crisis extension $e(R)=RLRRR...$ corresponds to the
last band merging point (where a critical point is mapped onto an
unstable fixed point after three iterations) and the sequence
$RLLL...$ to the final crisis point. All the windows of the tree
lie in between these two points (i.e.,  in the region of one chaotic band [1]).
\hfil\break
2) From two infinite "parent" sequences $A$ and $B$ form a finite
"daughter" sequence $C$ by taking the shared beginning of $A$ and $B$.
The first such sequence $RL$ is the "root" of the tree.\hfil\break
3) Take $A$ and $a(C)$ as the infinite parents of a new left "branch"
and $e(C)$ and $B$ as the infinite parents of a new right "branch".
Attach the symbol $0$ to the left and the symbol $1$ to the right branch.\hfil
\break
The beginning of the infinite binary tree is shown in Fig. 1.
There is a one-to-one correspondence between
the binary codes consisting of the symbols $0$ and $1$ and the MSS
sequences. In the following, we use the symbol $\to$ to express
this correspondence. For example, $1 0^2 \to RL^2 RLR$.
$\alpha^k$ means that the symbol (or a block of symbols) $\alpha$
is repeated $k$ times. This convention is used for both the MSS sequences
and the binary codes.

{\bf 3. TRANSFORMATION BETWEEN THE BINARY CODE AND THE
MSS SEQUENCE}

In this section we develop "self-contained" recursive rules by which the
transformation $\to$ can be carried out.
These rules are just another variant of the MSS rule but they
turn out to be the key to understanding how the
the lenght of the MSS sequence increases down the binary
tree. In the following,
the $i$th symbol of the MSS sequence $A$ is denoted by $\{ A\}_i$ and
the string from the $i$th symbol up to the $j$th symbol
by $\{ A\}_i^j$ ($j<i$ implies an empty string).
This notation is particularly useful if $A$ is an
extension or some other composition.

Assume that an infinite binary code $i_1 i_2 ...$ corresponds
to the MSS sequence $A$.
Let $A_k$ be the truncation of $A$ so that $i_1 i_2 ...i_k \to A_k$.
One of the "parent" branches of $A_{k+1}$ is always $A_k$.
The more distant parent of $A_{k+1}$
is denoted by $\hat A_{k+1}$ (and that of $A_k$ is $\hat A_k$).
For example, the parent branches of $A_3 =RL^2 RLR$ are
$A_2 =RL^2 R$ and $\hat A_3 =RL$.
$\hat A_k$ can be
defined also for the case in which $A_k$ lies at the edge of the binary
tree --- see below.
The infinite parents of $A_{k+1}$ are either $a(A_k )$ and
$e(\hat A_{k+1} )$ or $e(A_k )$ and $a(\hat A_{k+1} )$.
It is then clear that $A_{k+1}$ can be written either in the form
$$A_{k+1} =A_k \beta_k \{ a(A_k )\}_1^{h_k}$$
or
$$A_{k+1} =\hat A_{k+1} \gamma_{k+1} \{a(\hat A_{k+1} )\}_1^{m_{k+1}}$$
where $\beta_k$ and $\gamma_{k+1}$ are the first binding elements of
the proper extension.
$\beta_k$, $h_k$, and $m_k$ are needed in the construction of
the MSS sequence and can be determined by the following rules:

{\it Rule 1.} a) If $i_{k+1} =i_k$, then $\beta_k =
\{ a(\hat A_k )\}_{1+m_k}$
and $h_k$ is the length of the sequence
$$a(A_k )\wedge \{ a(\hat A_k )\}_{2+m_k}^{\infty}$$
b) $m_{k+1} =m_k +h_k +1$ and $\hat A_{k+1} =\hat A_k$.

{\it Proof.}
With $i_{k+1}=i_k$ the same parent branch is approached
as on the previous step so that $\hat A_{k+1} =\hat A_k$.
The immediate infinite parent of the new
daughter branch $A_{k+1}$ has the form $A_k \beta_k a(A_k )$, and
the more distant infinite parent has the form
$\hat A_k \gamma_k a(\hat A_k )$.
$A_{k+1}$ is their shared beginning. It can be longer than
$A_k =\hat A_k \gamma_k \{ a(\hat A_k )\}_1^{m_k}$
only if $\beta_k =\{ a(\hat A_k )\}_{1+m_k}$. In other words,
$$A_{k+1} =A_k \beta_k
[a(A_k )\wedge \{a(\hat A_k )\}_{2+m_k}^{\infty} ]$$
1b) can be easily verified by considering the lengths of
the sequences in the above two equations for $A_{k+1}$,
the latter also with the index $k$.
\hfill $\sqcap$

{\it Rule 2.} a) If $i_{k+1} \neq i_k$, then
$\beta_k =\{ a(A_{k-1} )\}_{1+h_{k-1}}$ and
$h_k$ is the length of the sequence
$$a(A_k )\wedge \{ a(A_{k-1} )\}_{2+h_{k-1}}^{\infty}$$
b) $m_{k+1} = h_{k-1} +h_k +1$ and
$\hat A_{k+1} =A_{k-1}$.

{\it Proof.} $i_{k+1} \neq i_k$ entails a turning back towards
the $k-1$'st branch. Thus, $\hat A_{k+1} =A_{k-1}$.
The infinite parents
of the new daughter branch have the forms
$A_{k-1} \beta_{k-1} a(A_{k-1} )$ and $A_k \beta_k a(A_k )$.
The shared beginning of the infinite parents can be longer than
$A_k =A_{k-1} \beta_{k-1} \{a(A_{k-1} )\}_1^{h_{k-1}}$
only if $\beta_k =\{ a( A_{k-1} )\}_{1+h_{k-1}}$. Thus,
$$A_{k+1} =A_k \beta_k
[a(A_k )\wedge \{a( A_{k-1} )\}_{2+h_{k-1}}^{\infty} ]$$
\hfill $\sqcap$

In fact, it would suffice to memorize $A_k$,
$h_{k-1}$, and $m_k$ because $\hat A_k$ and $A_{k-1}$
can be determined from these. If $i_1 =0$,
one has the initial conditions $A_1 =RLRR$, $h_0 =1$,
and $m_1 =2$ ($\hat A_1 =R $). If $i_1 =1$, then
$A_1 =RLL$, $h_0 =0$, and $m_1 =2$ ($\hat A_1 =\emptyset$).

{\bf 4. TYPICAL AND EXCEPTIONAL BINARY CODES}

The proofs of Rules 1 and 2 involve only a slight
elaboration on the MSS rule. The generation of the MSS sequence
with these rules is not necessarily
much more efficient than applying the
MSS rule directly. The new formulation helps
understanding why the length of the MSS sequence increases
arithmetically with some binary codes and geometrically with others.
The increasing length
of the MSS sequence in a single binary step is given by $h_k +1$.
We call the growth of the sequence length
arithmetic if $sup \{ H_1 ,H_2 ,...\} <H<\infty$, where
$$H_K ={1\over K}\sum_{k=1}^K h_k$$
This definition allows arbitrarily large occasional
increments in the MSS
sequence but they may not be frequent. If the growth is not
arithmetic it is called geometric.

$h_k$ must become large
for the MSS string length to grow rapidly. Let us denote
the length of $\hat A_{k+1}$ by $l(\hat A_{k+1} )$ and
assume that $m_{k+1} <l(\hat A_{k+1} )$
(call it the simple extension
condition). Then
both $A_k$ and $\hat A_k$ (Rule 1a) or $A_k$ and $A_{k-1}$ (Rule 2a)
are long enough so that we can replace the antiharmonic extensions with
the infinite MSS sequence $A$ in determining $h_k$.
This suggests that $h_k$ can be large
only if there is a large block in $A$
identical with the beginning of $A$ and which
lies after the first $1+m_k$
($i_{k+1} =i_k$) or $1+h_{k-1}$ ($i_{k+1} \neq i_k$) symbols of A.
On the other hand, the beginning of $A$ can be written as in the form
$A=\hat A_j \gamma_j \{a(\hat A_j )\}_1^{m_j }... $ anywhere
in the sequence (at arbitrary $j$). Note
that $a(\hat A_j )$ can be replaced by $A$ if
$l(\hat A_j )> m_j$. Thus $m_j$
gives a rough estimate of the length of the above described
block, identical with
the beginning of $A$, which comes following the first
$l(\hat A_j )+1$ symbols.
If
$m_k$ ($i_{k+1} =i_k$) or $h_{k-1}$ ($i_{k+1} \neq i_k$)
becomes equal to $l(\hat A_j )$ for some $j$ for which $m_j$ is
large, then $h_k$ can reach a large value.
If such an index $j$ is not found, then $h_k$ can be expected to
remain "small".

The initial values $h_0$ and $m_1$ are small, and Rule 2b implies
that taking steps in alternating directions in the binary tree
one is not likely to generate large values of $m_j$.
According to Rule 1b, $m_j$ grows at least linearly with $j$
if consecutive steps are taken in the same direction. In other words,
assuming that $I\bar i \; i^p \to \hat A_j \gamma_j \{ A\}_1^{m_j}$ with
$I\to \hat A_j$ we obtain $m_j \geq p$. In this way we see that
arbitrarily large values of $m_j$ and, accordingly, of $h_k$ are
possible. However, this is not sufficient to generate
geometric growth of an MSS sequence. Geometric growth requires
average unbounded increases of $h_k$ as a function of $k$.
It turns out that judiciously placed
blocks of identical symbols can bring about such a phenomenon.
It is difficult to explain the geometric growth of the MSS
sequence length for an arbitrary binary code.
In the following
we consider a special form of the binary code whose growth properties
are easier to understand.

{\it Proposition.} Consider binary codes of the form
$1010^{p(2)} 10^{p(3)} ... 10^{p(n)}$.
If $p(i)>0$ and $p(i)+j(i) <i$ ($i=2,3,...,n$),
where $j(i)$ is calculated from Eqs. (1-5) below,
then the corresponding MSS sequence has the form
$$\alpha_{-1} [\;\;]_{-1} \alpha_0 [\;\;]_0 ...\alpha_n [\;\;]_n$$
with $\alpha_n$ either $R$ or $L$ according to the
rule $\alpha_n = \alpha_{k(n)} $ ($k(n)<n$) beginning with
$\alpha_{-1}=R$ and $\alpha_0 =L$ and
$$
[\;\;]_n =[\;\;]_{k(n)} \alpha_{j(n)} [\;\;]_{j(n)}
\alpha_{1+j(n)} [\;\;]_{1+j(n)} ...\alpha_{m(n)} [\;\;]_{m(n)}
$$
Let $l(n)$ be the length of the block $[\;\;]_n$ and $s(n)$ the
length of the whole MSS sequence.
Then $k(n)$, $j(n)$, $m(n)$, $l(n)$, and $s(n)$ are given by
the recursion formulae
$$s(k(i)-1) =l(m(i-1)) \eqno(1)$$
$$s(j(i)-1) =l(k(i)) \eqno(2)$$
$$m(i)=j(i)+p(i)-1 \eqno(3)$$
$$l(i)=s(m(i)) \eqno(4)$$
$$s(i)=s(i-1)+l(i)+1 \eqno(5)$$
with the initial conditions $k(1)=0$, $m(1)=-1$, $l(-1)=l(0)=0$,
$l(1)=1$, $s(-2)=0$, $s(-1)=1$, $s(0)=2$, and $s(1)=4$.  The
MSS sequence begins with $RLL[R]_1 ...$

The proof is a straightforward application of Rules 1 and 2
and is omitted here. The condition
$p(i)+j(i) <i$ is equivalent to the simple extension condition.

Eqs. (1-4) can be combined into a recursion rule
for $m(n)$ alone:
$$m(n)=p(n)+m(m(m(n-1))+1) \eqno(6)$$
In addition to the initial condition for $m(1)$ one has to
specify the values $m(-1)=m(0)=-2$ in order to apply (6).
The simple extension condition in terms of $m(i)$ becomes
$m(i)< i-1$.

With the aid of Proposition binary sequences can be
constructed which
lead to geometric growth of the MSS sequence length.
Eqs. (4-5) give
$$s(n)=s(n-1)+s(m(n))+1 \eqno(7)$$
for the length of the sequence.
This implies that the window period grows according to
the recursion $q_n =q_{n-1} +q_{m(n)}$
with each added block $10^{p(n)}$ in the binary code.  The
powers $p(n)$ in Eq. (6)
can be chosen so that $m(n)=n-r$ with $r>1$ for
$n>N$.
The leading eigenvalue $\zeta$
of the transition matrix $M$ defining the recursion via
$(q_n ,q_{n-1},...,q_{n-r+1} )=M(q_{n-1} ,q_{n-2} ,...,q_{n-r})$
is greater than unity and gives
the asymptotic growth rate of the MSS sequence length.
The simple extension condition implies $\zeta <2$.

All Fibonacci sequences correspond to
$m(n)=n-2$ with $\zeta \approx 1.618$. Eq. (6) implies
that the power $p(n)$
takes the constant value $4$ after the "transients" have died
out.

{\it Example.} The MSS sequence for the
binary code $1010^2 10^3 10^4 10^4 ...$ reads
$$RLL[R]_1 R[RL]_2 R[RLL(R)]_3 L[RLL(R)R(RL)]_4
L[(R)LL(R)R(RL)R(RLLR)]_5 ...$$
Each block $\alpha_n [\;\;]_n$ results from a block $10^{p(n)}$
in the binary code. The
increments resulting from $h$ different from zero are
in parentheses (...).  These increments are
precisely the same as the blocks
$[\;\;]_i$ and $m(i)=i-2$ ($i=1,2,...$).

Examples of geometric growth with various $r$ and
asymptotic powers $p$
are listed in Table I.  All were
originally found numerically but the third, fourth and fifth
are beautifully explained by Proposition which predicts
correctly the binary period of families with the recursion
$q_n =q_{n-1} +q_{m(n)}$ although their binary
codes may not begin like 101....
More complicated cases can be constructed letting
$m(n)$ oscillate according to some rule.
For example, taking $m(2n+1)=2n-1$ and $m(2n)=2n-3$
leads to the recursion relations
$q_{2n+1} =q_{2n} +q_{2n-1}$ and $q_{2n} =q_{2n-1} +q_{2n-3}$ which
can be combined as $q_{2n+1} =2q_{2n-1} +q_{2n-3}$.
{}From Eq. (6) one can solve for the powers
$p(2)=1$, $p(3)=p(4)=3$, $p(5)=5$, $p(2n)=4$ ($n=3,4,...$), and
$p(2n+1)=6$ ($n=3,4,...$).  Each asymptotic binary block
$10^4 10^6$ means multiplying the period with the
average factor $\zeta \approx 1.554$.
Aperiodic binary codes can be generated making
the oscillations in $m(n)$ aperiodic.  Instead of letting
$m(n)$ oscillate regularly like above, we took
$m(n)=2[n/2]-3$ or $m(n)=2[n/2]-1$ at
random ($[\;\;]$ stands for the integer part),
beginning with an adjustable $n$.  The longer the
leading regular part, the faster the MSS sequences grew in
length.  On account of inherent limitations of computers,
no certain conclusion could be drawn about the asymptotic
algebraic or geometric growth of such sequences.  It may
be interesting to notice that the sum of the second differences
of the MSS lengths seldom seemed to display a sustained growth
suggesting an asymptotic algebraic fate to all MSS sequences
arising from random period binary codes.

A purely numerical method of finding geometric growth consists
of first defining a recursive rule for the increase of the
window period and then determining the  binary code which yields
the windows (only a small number of rules lead to simple behavior).
Each "appropriate" recursion rule has a characteristic repeating
pattern of the periodic binary tail.
For example the rule $q_n =q_{n-1} +q_{n-2}$
gives the same repeating pattern $10^4$ for all choices of
$q_1$ and $q_2$ (if one always moves to the right in the
binary tree).  Taking off from an arbitrary window, however,
and successively adding the pattern $10^4$ to the binary
code will almost certainly lead to asymptotically arithmetic
increase of the MSS sequence length!

Geometric growth obviously requires a very synchronous
binary code. A single mismatched binary symbol may suffice
to turn the growth into arithmetic. Therefore,
one expects the MSS sequence length to grow arithmetically
for most binary codes with periodic tails. We call such codes
"typical"  whereas the codes leading to the geometric growth
of the MSS sequence are called "exceptional".

The examples of Table I have been constructed by
always picking the closest matching window on the right
hand side.  We could not find
any exceptional codes with periodic tails by choosing
the next window on the left hand side.  One may begin, nevertheless,
with a binary code that carries a number of leading zeros.

{\bf 5. NUMERICAL SCALING RESULTS}

The scaling behavior is strongly related to the manner of
growth of the length of the MSS sequence along
a path in the binary tree. In the following, the scaling
of both typical and exceptional binary codes is discussed
concentrating on codes with periodic tails.

{\bf 5.1 Typical codes with periodic tails}

The binary sequence $1^{\infty}$ corresponds to the stable periods
$4,5,6,...$
approaching the final crisis point. The scaling properties of this
sequence are well understood [10,11]. Along the parameter axis
the scaling of the superstable parameter values
is geometric. The scaling factor is determined by the
derivative of the map at the unstable fixed point which is the image
of the critical point under the second iterate of the map. The widths
of the windows scale by the square of this factor.
Every other binary code with the tail $1^{\infty}$ corresponds to
a window sequence approaching a tangent bifurcation point, i.e. the
left end point of some periodic window.
It has been shown [12] that this results in slower than
geometrical scaling.

On the other hand, every binary sequence with
the tail $0^{\infty}$ leads to a sequence which accumulates at an
internal crisis point. All these points are fully chaotic according
to a theorem by Misiurewicz [1,4]. In the same way as for the final
crisis point, the scaling properties are determined by the Lyapunov
factors of the associated unstable orbits (Section 6).

Let us now consider an arbitrary binary code of
the form $IJ^{\infty}$ and let $\mu_k$ and $\Delta \mu_k$ be the
superstable parameter value of the window $IJ^k$
and its width. We determine
the scaling factors
$$\delta_k ={\mu_k -\mu_{k-1} \over \mu_{k+1} -\mu_k};\;\;\;
\sigma_k ={\Delta \mu_{k-1} \over \Delta \mu_k } $$
for the logistic map
and find the period $\nu$ by which $\delta_k$ and
$\sigma_k$ oscillate as $k\to \infty$.
The asymptotic limits of the products of $\nu$ subsequent
$\delta_k$'s and $\sigma_k$'s are denoted by $\delta$ and $\sigma$,
respectively.  Table II displays $\nu$
and $\delta$ for a number of
examples. Neither $\nu$ nor $\delta$
is universally determined by the tail
of the binary code. In Section 6 it is shown that both
are related to the orbit of the critical point at the accumulation
point of the window sequence.  The accumulation point
for this class of codes is always a Misiurewicz point
at which the critical point is mapped onto an unstable period
$M_p$ after $M_i$ iterations.
A proof of this claim is given in Appendix. The numbers
$M_p$ and $M_i$ are included in Table II.
It turns out that an MSS sequence corresponding to the typical
binary code has a periodic tail,
which is possible only if the critical point is mapped onto an
unstable orbit. The repeating pattern gives the kneading sequence
of the unstable orbit. The quantity
$\nu$ is greater than unity if the whole pattern
is not traversed in one period of the binary tail.
It may happen that the pattern is run through more
than once per binary period.
In Table II, $\rho$ gives the number of times the pattern is
completed in $\nu$ periods of the binary tail.

In Section 6 the result of Post and Capel [11] that
$\sigma =\delta^2$ is generalized to all codes within this class.
In other words, the scaling behavior in this class is equivalent
to the one for the code $1^{\infty}$ and those with the tail
$0^{\infty}$.

{\bf 5.2 Exceptional codes with periodic tails}

It has been shown in the Fibonacci case that
the scaling of the superstable parameter values is
superexponential [13,14].  A geometric increase
of the MSS sequence length leads to  superexponential scaling
in all the cases we studied.
Furthermore, a comparison of the appropriate scaling factors [13,14]
for the logistic map on the one hand and
$f_\mu (x)=\mu \sin (\pi x)$ on the other suggests that the
superexponential scaling is universal.

{\bf 5.3 Codes with aperiodic tails}

It was demonstrated in Section 4 that
exceptional aperiodic codes with rapid growth rates of
the MSS sequence length can be explicitly constructed.
For a random code
with no build-in blocks of identical symbols,
however, it is natural to expect arithmetic average growth.
The numerical studies bear out this expectation.
We observe average geometric scalings of the superstable
parameter values and positive Lyapunov exponents at
the accumulation points.
\vfill\eject

{\bf 6. GEOMETRIC SCALING}

We are interested in families of MSS
sequences whose binary codes have periodic tails.  These
sequences have some initial length $M_i$.  They
grow in steps of $n=\rho M_p$ per advancing $\nu$ periodic blocks
of the binary tail.
In order to get at the scaling parameters $\delta$ and
$\sigma$ defined in the preceding section,
we take advantage of the local description
of Post and Capel [11]. It describes the map $f_\mu^l (x)$
within the window of period $l$.
We choose a particular representation
of the logistic map $f_\mu (x)=\mu - |x|^z$ where $z$ may
take the values $1< z< \infty$ (i.e., in this section we
do not restrict ourselves to the case of a quadratic maximum). Then the local
descriptions is centered at $x=0$, and the following
reduced form applies
$$
x_{l(i+1)}=\rho_l (\mu ) +\lambda_l |x_{li} |^z +
\kappa_l |x_{li} |^{2z} +... \eqno(8)
$$
where $x_{l(i=0)}$ is some starting point close to the central
value $x=0$, so close that we do not have to worry about
the term to the power $2z$ on the right hand side. The local
description then incorporates one parameter $\lambda_l $.
Finally
$$\rho_l (\mu )=\mu -|\mu -|\mu -...|\mu|^z|^z|^z...|^z \eqno(
9) $$
with $l-1$ pairs of vertical bars.
For the present purposes the index $l$ will need to take
values like $m+kn$ and $m+(k+1)n$ where $k$ is a large integer
since we are interested in the asymptotic $\mu$ scaling and the
ratios of widths of consecutive
windows in our families.

With the substitutions
$$
x_{li}=u_i |\lambda |^{-1/(z-1)}sgn\lambda
$$
$$
\rho =-r|\lambda |^{-1/(z-1)}sgn\lambda \eqno(10)
$$
and leaving out the subscript of $\lambda$,
the above form turns into Post's and Capel's normalized
submap
$$u_{i+1}=|u_i|^z-r \eqno(11) $$

First consider the ratios of window widths.  If $\rho$
varies by $\Delta\rho$ when
$\mu$ varies across the period-$l$ window, the window width expressed
in $\mu$ is obviously
$$ \Delta\mu \simeq \Delta\rho
{\bigr (
{
{{d\rho} \over {d\mu }} \bigr |_{\mu =\tilde \mu_l }
} \bigr )
}^{-1}$$
where a tilde on the symbol $\mu$ indicates the superstable
parameter value.
By the scaling, Eq. (10), between $\rho$ and $r$ we can express
$\Delta\rho$ in terms of $\Delta r$, the invariable normalized window
width of Eq. (11).
The "physical" window width $w_l $ is then
$$ w_l = \Delta\mu \simeq {
{\Delta r} \over
{ |\lambda |^{1/(z-1) } \bigr (
{{d\rho} \over {d\mu }} \bigr |_{\mu=\tilde \mu}
} \bigr )
}$$
We now give the index $l$ the values $N+n$ and $N$ where
$N=m+kn$.
For ratios of window widths, $\Delta r$ cancels, and the scaling
in the family becomes asymptotically
$$ \hat \sigma_k=w_N /w_{N+n} =
{ {
\bigr (
{
{\lambda_{N+n}}
\over {\lambda_N }
}
\bigr )
}^{1/(z-1)}}
{{{{d\rho} \over {d\mu }}\bigr |_{\mu =\tilde \mu_{N+n} }
}
\over
{ {{d\rho } \over {d\mu }} \bigr |_{\mu =\tilde \mu_N}
}}  \eqno(12)
$$
when $N=m+nk$ becomes large, i.e. $k$ becomes large
($\hat \sigma_k$ is essentially the product of $\nu$ subsequent
$\sigma_k$'s defined in Section 5).

For the scaling of the positions of the windows we turn to the
function $\rho_{m+kn}$.  It determines the height of the
quadratic maximum or minimum of the local mapping $f_\mu^{m+kn} (x)$
and it gives
the value of $x$ to which the center point $x=0$ is sent
in $f_\mu^{m+kn} (x)$.  At $\mu=\tilde \mu_{m+nk}$,
$\rho_{m+nk} =\rho (\mu )$ takes the value zero.  Let us look at the
functions $\rho_{m+nk}$ for different $k$ at the accumulation
point $\mu_{\infty }$ of the family.  Denote again
$N=m+nk$.  From the definition of $\rho$ above, Eq. (9)
$$
\rho_N (\mu ) =f^N_\mu (0)
$$
and
$$
\rho_{N+n} (\mu )=f^n_\mu ( \rho_N)
$$
But the orbit of the center $x=0$ repeats itself with
the period $n$. It follows
$$
\rho_{N+n} (\mu_{\infty }) \simeq \rho_N (\mu_{\infty })
$$
which means that the $\rho_{m+kn}$ for different $k$ all
meet at the same point at $\mu_{\infty }$. Assuming that the
$\rho (\mu )$  are locally straight lines, we get the asymptotic
$\mu$ scaling from the
derivatives of the $\rho$ for different $k$.  Estimate
$\rho_N$ and $\rho_{N+n}$ as follows
$$
\rho_N (\mu_{\infty }) \simeq - {{d\rho_N } \over {d\mu }} \bigr
|_{\mu = \tilde \mu_N } (\tilde \mu_N -\mu_{\infty })
$$
and
$$
\rho_{N+n} (\mu_{\infty })\simeq -{{d\rho_{N+n} } \over {d\mu}}
\bigr |_{\mu = \tilde \mu_{N+n}} (\tilde \mu_{N+n}-\mu_{\infty })
$$
These two quantities being equal,
$$
{{\tilde \mu_N -\mu_ {\infty }} \over {\tilde \mu_{N+n} -\mu_{\infty }}}
\simeq {{{d\rho_{N+n} } \over {d\mu }} \bigr |_{\mu = \tilde \mu_{N+n}} \over
{{{d\rho_N } \over {d\mu }} \bigr |_{\mu = \tilde \mu_N }}}
$$
and
$$
{\hat \delta_k ={{\tilde \mu_N-\tilde \mu_{N+n} }
 \over {\tilde \mu_{N+n} -\tilde \mu_{N+2n}}}\simeq
{{\tilde \mu_N -\mu_ {\infty }} \over {\tilde \mu_{N+n}
-\mu_{\infty }}}
={\bigr ( {{d\rho_{N+n} } \over {d\mu }} \bigr |_{\mu =
\tilde \mu_{N+n}}} \bigr ) }/ {
\bigr ( {{d\rho_N } \over {d\mu }} \bigr |_
{\mu = \tilde \mu_N} \bigr )} \eqno(13)
$$
again for large $N$.
We have not seen this result in the literature.

What remains is calculating the scaling factors $\lambda$
and the ratio of $(d\rho_{N+n} /d\mu)$ at $\mu =\tilde \mu_{N+n}$
and $d\rho_N /d\mu $
at $\mu =\tilde \mu_N $ .  We follow Post
and Capel [11].

Take $\lambda $ first.  In the following, a tilde on the
variable $x$ refers to an iterate of the central value
$x=0$.
At superstability, since $\rho$ vanishes there,
we have from the local description of Eq. (8)
$$
x_{l(i+1)}=\lambda |x_{li} |^z \eqno(14)
$$
where some starting point $x_{li}$ close to the central value
$x=0$
has been picked.  We will express the left hand side of
this equation as an expansion in terms of the following
quantity
$$
v=f_{\tilde \mu_l }(x_{li} )-f_{\tilde \mu_l }(\tilde x_0)
$$
with the same point $x_{li}$ .
In a single shot
of the mapping $f_{\tilde \mu_l}$, the central point and
a point in its vicinity will be sent far away from the center
but roughly to the same location.  Therefore the quantity $v$ is
small.  To linear order in $v$
then
$$
x_{l(i+1)}=f_{\tilde \mu_l}^{l-1}(f_{\tilde \mu_l}(x_{li}))
=f_{\tilde \mu_l}^{l-1}(f_{\tilde \mu_l}(\tilde x_0)+v)
$$
$$
=f_{\tilde \mu_l}^{l-1}(f_{\tilde \mu_l}(\tilde x_0))
+{f_{\tilde \mu_l}^{(l-1)}}^\prime (f_{\tilde \mu_l}(\tilde x_0))v
=\prod_{j=1}^{l-1} {f_{\tilde \mu_l}}^\prime (\tilde x_j)v \eqno(15)
$$
where a prime denotes a derivative with respect to the
argument of the function at the indicated value of the
argument.
For the specific form $ f(x)=\mu-|x|^z$
$$
v=-|x_{li}|^z
$$
and Eq. (15) yields
$$
x_{l(i+1)}=- \prod_{j=1}^{l-1} {f_{\tilde \mu_l}}^\prime (\tilde x_j)
|x_{li}|^z
$$
and one reads from Eq. (14)
$$
\lambda=-\prod_{j=1}^{l-1}{f_{\tilde \mu_l}}^\prime (\tilde x_j)
\eqno(16)
$$

With $\lambda_l$ cleared, one still needs to
calculate the derivative of $\rho$ with
respect to $\mu$ in Eqs. (12) and (13).
Look again at the reduced map $f_{\mu}^l (x)$
$$
x_l=\rho + \lambda |x_0|^z
$$
and its derivative with respect to $\mu$
$$
{{dx_l} \over {d\mu}}={{d\rho} \over {d\mu}}+{d \over {d\mu}}
\lambda |x_0|^z
$$
Choose $x_0$ as $\tilde x_0=0$ and take the expression
at the superstable point.
The second term on the right vanishes and
$$
{{d\tilde x_l} \over {d\mu}} \bigr|_{\tilde \mu_l} ={{d\rho} \over {
d\mu}} \bigr|_{\tilde \mu_l}
$$
One thus needs an expression for $({d\tilde x_l/d\mu)}_{\tilde \mu_l}$.
Consider the map
$$
x_{i+1}=f_{\mu} (x_i)= \mu -|x_i|^z
$$
In general $x_{i+1}$ depends on $x_0$ in addition to
$\mu$ which is indicated by the partial derivatives in the
sequel
$$
{{\partial x_{i+1}} \over {\partial \mu}}=1-z|x_i|^{z-1} sgn x_i
{{\partial x_i} \over {\partial \mu}}=
1+{f_{\mu}}^\prime (x_i) {{\partial x_i} \over {d\mu }}
$$
$$
=1+{f_\mu}^\prime (x_i )[1+{f_{\mu}}^\prime
(x_{i-1})[1+{f_{\mu}}^\prime (x_{i-2})
[1+\cdots [1+{f_{\mu}}^\prime (x_0)]]]\cdots ] \eqno(17)
$$
Nothing prevents us from choosing i+1=l and picking the
superstable $\mu_l$ along with $x_0=\tilde x_0=0$.  Then the
iterates depend only on $\mu$ and
$$
{{d\tilde x_l} \over {d\mu}} \bigr |_{\tilde \mu_l}=
1+{f_{\tilde \mu_l}}^\prime (\tilde x_{l-1}) [1+
{f_{\tilde \mu_l}}^\prime (\tilde x_{l-2}) [1+ \cdots
[1+{f_{\tilde \mu_l}}^\prime (\tilde x_1)]]\cdots ]=
1+\sum_{i=1}^{l-1} \prod_{j=i}^{l-1} {f_{\tilde \mu_l}}^\prime
(\tilde x_j )
$$
or finally
$$
{{d\rho} \over {d\mu}} \bigr |_{\tilde \mu_l}=
1+\sum_{i=1}^{l-1} \prod_{j=i}^{l-1} {f_{\tilde \mu_l}}^\prime
(\tilde x_j)
$$

Now return to Eq. (12).  From Eq. (16) we get
immediately at the limit when $N \to \infty $
$$
\lambda_{N+n}/\lambda_N=\prod_{i=K}^{K+n} f_{\tilde \mu_{N+n}}^\prime
(\tilde x_i)
$$
where K is some number larger than $M_i$, the number of initial
iterations before hitting the unstable period, and smaller
than $N-M_f$, $M_f$ being the number of the last
steps drifting away from the unstable period to end
at the center.  It is important to notice for what follows
that $K$ may vary over the whole middle range of the $N+n$
cycle.
It turns out convenient to call $L_i$ the derivatives
in the product and $L$ the whole product from $i=K$
to $i=K+n$.
The contribution to $\hat \sigma_k $ of the first factor
in Eq. (12) is then
$$
{\bigr ( {{\lambda_{N+n}} \over {\lambda_N}} \bigr)}^{1/(z-1)}
=L^{1/(z-1)} \eqno(18)
$$
The second quantity in Eq. (12),
$$
\bigr ( {{d\rho} \over {d\mu}} \bigr |_{\mu=\tilde \mu_{N+n}}
\bigr )/\bigr ( {{d\rho} \over {d\mu}} \bigr |_{\mu=\tilde
\mu_N} \bigr ) \eqno(19)
$$
has a structure which is easiest grasped looking at one
derivative at a time.  In the following the products are written
in the order of increasing length, more or less like in
Eq. (17).  We need clarifying notation:
$$
1+L_{N-1}+L_{N-1}L_{N-2}+L_{N-1}L_{N-2}L_{N-3}+ \cdots +e=E
$$
where $e=L_{N-1}L_{N-2}L_{N-3}\cdots L_{N- M_f}$.
Then
$$
{{d\rho} \over {d\mu}} \bigr |_{\mu=\tilde \mu_N}
=1+L_{N-1}+L_{N-1}L_{N-2}+L_{N-1}L_{N-2}L_{N-3}+\cdots +L_{N-1}
L_{N-2}\cdots L_1
$$
$$
=E+e(L_{N-M_f-1}+L_{N-M_f-1}L_{n-M_f-2}+\cdots
$$ $$ +L_{N-M_f-1}
L_{N-M_f-2}\cdots L_{N-M_f-n+1}+L)
$$
$$
+eL(L_{N-M_f-n-1}+
L_{N-M_f-n-1}L_{N-M_f-n-2}+\cdots
$$ $$ +
L_{N-M_f -n-1}L_{N-M_f -n-2}\cdots L_{N-M_f-2n+1}
+L)+\cdots  \eqno(20)
$$
Now the expressions in the brackets in Eq. (20) are
identical since they run over full unstable periods of length $n$.
We again
introduce a new symbol $B$, this time for the sum in the
brackets multiplied by $e$.  Then
$$
{{d\rho} \over {d\mu}} \bigr |_{\mu=\tilde \mu_N}=
E+B+BL+BLL+\cdots + BL^{(N-M_f-M_i)/n}+P \eqno(21)
$$
where
$$
P=eL^{(N-M_f-M_i)/n}(L_{M_i-1}+L_{M_i-1}
L_{M_i-2}+\cdots +L_{M_i-1}\cdots L_1 )
$$
The quantity $M_f$ may be thought of as having been chosen
such that $(N-M_f-M_i)/n$ is an integer.  $P$ represents the
in-going steps before hitting the unstable period.

Introduce yet another symbol $S=S(N)$ for the right hand
side of Eq. (21) without the term $E$.
Remember that the ratio we are seeking to calculate,
Eq. (19), is between two expressions of type
Eq. (21) with the numerator having $N+n$
in the place of the denominator's $N$.  If the denominator
is expressed as in Eq. (21), the numerator
cycle is longer by $n$ and the right hand side of Eq. (21)
will have one more term,
$$
BL^{(N+n-M_f-M_i)/n}
$$
and the factor $P$ becomes multiplied with $L$.
This can be interpreted as multiplying by $L$ the terms which were
given the name $S$ above and adding back a $B$. It follows
$$
\bigr ( {{d\rho} \over {d\mu}} \bigr |_{\mu=\tilde \mu_{N+n}}
\bigr )/\bigr ( {{d\rho} \over {d\mu}} \bigr |_{\mu=\tilde
\mu_N} \bigr ) =
{{E+B+LS} \over {E+S}} \eqno(22)
$$
$S$ is obviously rapidly growing function of $N$ provided that $L$
is larger than unity, i.e. that the periodic orbit at the accumulation
point of the family is unstable.
All other quantities in Eq. (22) remain
constant when $N$ grows. The asymptotic result at the limit
$N \to \infty$ is simply
$$
\bigr ( {{d\rho} \over {d\mu}} \bigr |_{\mu=\tilde \mu_{N+n}}
\bigr )/\bigr ( {{d\rho} \over {d\mu}} \bigr |_{\mu=\tilde
\mu_N} \bigr ) =L
$$
By Eq. (13) this result is directly the parameter scaling
factor $\delta$.
$$
\delta =L
$$
Together with Eq. (18) we get
for the scaling of the window width, Eq. (12)
$$
\sigma =L^{1/(z-1)}L=L^{z/(z-1)}
$$

{\bf 7. DISCUSSION}

It was shown in Section 6 that
typical binary codes with periodic tails lead to geometric
scaling whatever
the order $z$ of the critical point.
The situation is very different in the case of the exceptional codes.
Our numerical results suggest (see Section 5)
that the scaling for an exceptional code with periodic tail
 is always superexponential
in the case of the logistic
map with a quadratic maximum.
If we let the order of the maximum increase, however,
the scaling becomes
geometric and  universal. For the Fibonacci sequences this has been
pointed out by M. Lyubich [19]. The critical value of the exponent
$z$ for the Fibonacci sequences is $2$. In other cases it is
numerically much harder to pinpoint the exact critical value
of the universality class at which the transition from
superexponential to geometric scaling takes place. This would be
an interesting problem for future studies.

A binary tree of stable periodic
attractors appears also in invertible circle maps [20]. It is
therefore
interesting to compare the scaling behaviors in these two cases.
In a circle map the period of the attractor increases geometrically
for almost all routes in the binary Farey tree. In the standard case
this leads to universal geometric scaling. In the unimodal map a
geometric increase of the period
and the resulting universal scaling is observed only in some
exceptional cases. It is an open question whether one could construct
a "non-renormalizable" circle map with superexponential
scaling. It is not clear how the various scaling
properties relate to the Lebesgue measure of the parameter values
for aperiodic
attractors. For a standard
critical cubic circle map this measure is zero whereas
for the quadratic unimodal map the
measure of aperiodic attractors is positive.

The typical binary codes with periodic tails give a huge number of new
Misiurewicz points in a unimodal map (recall that the same structure is
repeated within each window). These points form a subset of
the set of all Misiurewicz points
of the Mandelbrot set [21]. It would be an interesting
problem in the theory of numbers
to work out whether the typical binary codes with periodic
tails (taking into account the self-similarity) give all
the Misiurewicz points of a unimodal map.

{\bf APPENDIX. A TYPICAL BINARY CODE WITH A PERIODIC TAIL LEADS TO
A MISIUREWICZ POINT}

Let us assume that the binary code has a periodic tail which
is neither $0^{\infty}$ nor $1^{\infty}$. The code
can always be written in the form $IJJJ...$, where
$I$ and $J=j_1 ...j_K$ are finite binary
codes with $j_2 =\bar j_1$.
Let $A$ be the corresponding infinite MSS sequence.

{\it Lemma.} Consider a binary code of the above form with
$h_k <H<\infty$ for every $k$.
Then the accumulation point of the corresponding sequence
of periodic windows is a Misiurewicz point.

{\it Proof.} Because $j_2 =\bar j_1$, Rule 2b implies that
$m_{N+2+qK} <2H$ for $q=0,1,2,...$, where $N$ is the length of $I$.
Furthermore, by Rules 1b and 2b $m_{k+1} \leq m_k +H$.
Combining these two results gives an upper bound for $m_k$:
$m_k <(K+1)H$ for $k=N+2,N+3,...$. Thus, there exists
a $p\geq N+2$ such that $m_p <l(\hat A_p )$.
Because $m_{k+1} < l(\hat A_{k+1} )$ for $k=p-1,p,...$,  it is
not necessary to carry out the antiharmonic extensions when
determining $h_k$ ($k\geq p-1$) in Rules 1a or 2a. In particular,
$A$ can be written as the composition
$A_{p-1} \beta_{p-1} \{ A\}_1^{h_{p-1}} \beta_p
\{ A\}_1^{h_p} ...$. The set of possible values for $m_k$ and $h_k$
is finite when $k>N+1$. Therefore,
there exist finite $P>p$ and $Q$ so that
$(m_P ,h_P ,h_{P-1} )=(m_{P+QK} ,h_{P+QK} ,h_{P+QK-1} )$.
Because $i_{P+k} =i_{P+QK+k}$ for $k=0,1,2...$, Rules 1 and 2
imply that
$(m_{P+k} ,h_{P+k} ,h_{P+k-1} )=(m_{P+QK+k} ,h_{P+QK+k} ,h_{P+QK+k-1} )$
which is possible only if $A$ has a periodic tail.
By Corollary II.8.4 of ref. [1], $A$ is an MSS sequence for a
Misiurewicz point.
\hfill $\sqcap$

\vfill\eject

{\bf REFERENCES}

[1] P. Collet and J.-P. Eckmann, Iterated maps on the interval as dynamical
systems (Birkh\"auser, Basel, 1980).

[2] M. Metropolis, M.L. Stein, and P.R. Stein, J. Combinatorial Theory
A 15, 25 (1973).

[3] J. Milnor and W. Thurston, in Dynamical systems, Proc. U. Md.,
1986-87, ed. J. Alexander, Lect. Notes Math. 1342 (Springer, 1988)
p. 465.

[4] M. Misiurewicz, Inst. Hautes \'Etudes Sci. Publ. Math.
53, 17 (1981); for the most recent developments see
G. Keller and T. Nowicki, Commun. Math. Phys. 149, 31 (1992),
and references therein.

[5] M. Jakobson, Commun. Math. Phys. 81, 39 (1981);
M. Benedicks and L. Carleson, Ann. Math. 122, 1 (1985).

[6] M.J. Feigenbaum, J. Stat. Phys. 21, 669 (1979).

[7] J.A. Ketoja and J. Kurkij\"arvi, Phys. Rev. A 33, 2846 (1986).

[8] J.D. Farmer, Phys. Rev. Lett. 55, 351 (1985);
C. Grebogi, S.W. McDonald, E. Ott, and J.A. Yorke, Phys. Lett. A 110,
1 (1985); G. Gao and G. Hu, Commun. Theor. Phys. 10, 127 (1988).

[9] J.-P. Eckmann, H. Epstein, and P. Wittwer, Commun. Math. Phys. 93, 495
(1984); R. Delbourgo and B.G. Kenny, Phys. Rev. A 33, 3292 (1986).

[10] T. Geisel and J. Nierwetberg, Phys. Rev. Lett. 47, 975 (1981);
V. Urumov and L. Kocarev, Phys. Lett. A 144, 220 (1990).

[11] T. Post and H.W. Capel, Physica A 178, 62 (1991).

[12] J. Dias de Deus, R. Dil\~ao, and A. Noronha da Costa, Phys. Lett. A 101,
459 (1984).

[13] K. Shibayama, in: The theory of dynamical systems and its
applications to nonlinear problems,
ed. H. Kawakami (World Scientific, Singapore, 1984) p. 124.

[14] J.A. Ketoja and O.-P. Piiril\"a, Phys. Lett. A 138, 488 (1989).

[15] M. Lyubich and J. Milnor, The Fibonacci unimodal map.
Preprint IMS91-15.

[16] Y. Ge, E. Rusjan, and P. Zweifel, J. Stat. Phys. 59, 1265 (1990).

[17] B. Derrida, A. Gervois, and Y. Pomeau, J. Phys. A 12, 269 (1979).

[18] M.J. Feigenbaum, J. Stat. Phys. 52, 527 (1988).

[19] M. Lyubich, A talk at the Workshop on Renormalisation in Dynamical
Systems, University of Warwick, Coventry, 30 March 1992.

[20] P. Cvitanovi\'c, B. Shraiman, and B. S\"oderberg, Physica Scr.
32, 263 (1985).

[21] H.-O. Peitgen and P.H. Richter, The Beauty of Fractals
(Springer, Berlin, 1986).

\vfill\eject

Table I.
The repeating patterns of the periodic binary tails for some
recursive rules leading to
an asymptotic geometric increase (given by the factor $\zeta$)
of the MSS sequence length.

$$\vbox{\offinterlineskip
\hrule
\halign{&\vrule #&\strut\quad \hfil# \quad\cr
&rule&&$\zeta$&&pattern&\cr
\noalign{\hrule}
&$q_n =2q_{n-1} +q_{n-2} $&&$2.414$&&$10^{10}$&\cr
&$q_n =q_{n-1} +2q_{n-2} $&&$2.000$&&$10^3$&\cr
&$q_n =q_{n-1} +q_{n-2} $&&$1.618$&&$10^4$&\cr
&$q_n =q_{n-1} +q_{n-3} $&&$1.466$&&$10^6$&\cr
&$q_n =q_{n-1} +q_{n-4} $&&$1.380$&&$10^8$&\cr
&$q_n =q_{n-2} +q_{n-3} $&&$1.325$&&$10^{10}$&\cr
&$ $&&$ $&&$ $&\cr}
\hrule}$$

Table II. The scaling period $\nu$ and the factor $\delta$
for some typical binary codes with periodic tails.
$M_p$ gives the period of the unstable orbit and
$M_i$ the length of the transient at the accumulation point.
The repeating MSS pattern is gone through $\rho$ times
during $\nu$ periods of the binary tail.

$$\vbox{\offinterlineskip
\hrule
\halign{&\vrule #&\strut\quad \hfil# \quad\cr
&code&&$\nu$&&$\rho$&&$\delta$&&$M_p$&&$M_i$&\cr
\noalign{\hrule}
&$(01)^{\infty}$&&$1$&&$2$&&$6.996$&&$2$&&$4$&\cr
&$(10)^{\infty}$&&$1$&&$2$&&$3.716$&&$1$&&$4$&\cr
&$0(01)^{\infty}$&&$1$&&$2$&&$5.560$&&$2$&&$6$&\cr
&$1(10)^{\infty}$&&$1$&&$2$&&$3.931$&&$1$&&$5$&\cr
&$(001)^{\infty}$&&$1$&&$1$&&$12.11$&&$6$&&$4$&\cr
&$(010)^{\infty}$&&$2$&&$1$&&$53.96$&&$9$&&$10$&\cr
&$(100)^{\infty}$&&$1$&&$1$&&$4.962$&&$3$&&$8$&\cr
&$(110)^{\infty}$&&$1$&&$1$&&$7.694$&&$3$&&$4$&\cr
&$(101)^{\infty}$&&$1$&&$1$&&$6.736$&&$3$&&$3$&\cr
&$(011)^{\infty}$&&$2$&&$1$&&$95.30$&&$9$&&$4$&\cr
&$ $&&$ $&&$ $&&$ $&&$ $&&$ $&\cr}
\hrule}$$

\vfill\eject

{\bf FIGURE CAPTION}

Fig. 1. The beginning of the infinite binary tree of periodic
windows.

\vskip 2cm

\line{\hskip 6.8cm $RL$ \hfil}
\bigskip

\line{\hskip 2.83cm $RLR^2$ \hskip 5.9cm $RL^2$ \hfil}
\bigskip

\line{\hskip 1.1cm $RLR^4$ \hskip 2.2cm $RLR^2LR$ \hskip 2.2cm $RL^2 R$
\hskip 2.10cm $RL^3$ \hfil}
\bigskip

\line{$RLR^6$ \hskip 0.3cm $RLR^4LR$ \hskip 0.3cm $RLR^2 LRLR$
\hskip 0.2cm $RLR^2 LR^2$ \hskip 0.2cm $RL^2 RLR$ \hskip 0.2cm
$RL^2 R^2$ \hskip 0.3cm $RL^3 R$ \hskip 0.30cm $RL^4$ \hfil}

\end